# Mechanosensitive Ion Permeation across Sub-Nanoporous MoS$_2$ Monolayers

A. Fang, K. Kroenlein, and A. Smolyanitsky*

Applied Chemicals and Materials Division,
National Institute of Standards and Technology
Boulder, CO 80305

*To whom correspondence should be addressed: alex.smolyanitsky@nist.gov

**Abstract**

We use all-atom molecular dynamics simulations informed by density functional theory calculations to investigate aqueous ion transport across sub-nanoporous monolayer molybdenum disulfide (MoS$_2$) membranes subject to varying tensile strains. Driven by a transmembrane electric field, highly mechanosensitive permeation of both anions and cations is demonstrated in membranes featuring certain pore structures. For pores that are permeable when unstrained, we demonstrate ion current modulation by a factor of over 20 in the tensile strain range of 0 – 4%. For unstrained pores that are impermeable, a clear strain-induced onset of permeability is demonstrated within the same range of strains. The underlying mechanism is shown to be a strain-induced reduction of the generally repulsive ion-pore interactions resulting from the ions' short-range interactions with the atoms in the pore interior and desolvation effects. The mechanosensitive pores considered in this work gain their electrostatic properties from the pore geometries and in principle do not require additional chemical functionalization. Here we propose the possibility of a new class of mechanosensitive nanoporous materials with permeation properties determined by the targeted engineering of vacancy defects.

**Introduction**

Systems involving permeation of ions across a porous membrane in aqueous solution have a variety of applications, ranging from water desalination to chemical sensing.[1] Nanoporous two-dimensional (2D) membranes are especially promising as materials for selective ion transport due to their advantages of high permeance, mechanical strength, and ability to in principle allow tailoring of pore properties via modifying pore size or chemical functionalization.[1] A property of particular interest is the ability to actively modulate the amount of ion permeation. This is highly desirable for numerous applications in nanofluidics[2] and is also a defining feature of biological ion channels, which regulate salinity within cells for a host of biological processes. Control of ionic flow via artificial nanopores can be achieved by for example voltage[3] or pH[4] gating. In addition, graphene-embedded crown ether pores have been recently predicted to exhibit cation permeation that is highly sensitive to applied tensile strain.[5] However, fabrication of such pores effectively requires functionalization of the pore lining with oxygens or partial reduction of graphene oxide.[6] Furthermore, until now there have been no demonstrations of membranes that exhibit strain-tunable anion permeation.

Single-layer $MoS_2$ is rapidly emerging as a highly promising 2D material for porous membranes due to its thickness of less than 1 nm, demonstrated stability when subjected to a transmembrane voltage in aqueous ionic solution,[7] and relative hydrophilicity compared to graphene.[8-9] Various types of nanopores in $MoS_2$ have been fabricated experimentally using electron[7] and ion[10] irradiation or by initiating electrochemical reactions at defects,[11] motivated by applications such as DNA sensing,[7] fundamental studies of ionic transport,[12] and "blue energy" harvesting.[13] One potential advantage of using $MoS_2$ for nanoporous membranes is that in principle no chemical functionalization is needed to create pores featuring desired electrostatics at the pore interior, since targeted atom removal may be a possible strategy for obtaining pore linings with engineered electrostatic environments at the pore interior due to the different electronegativities of Mo and S atoms. Several examples of ultra-narrow pore configurations in a hexagonal $MoS_2$ monolayer are shown in Fig. 1. Quantum-chemical calculations show that the Mo and S atoms carry partial electrostatic charges of opposite sign (for example, $q_{Mo}$ = +0.5e and $q_S$ = -0.25e in bulk $MoS_2$ monolayers[9]). Thus, radially dipolar environments are expected at the interiors of the pores in Figs. 1c-h, in addition to a possible effective total charge that may be carried by the pore. In principle, varying the radial dipole polarity and thus the permeation selectivity is possible by tuning the pore geometry (compare e.g., Figs. 1e and f, or 1g and h), unlike in the case of graphene-embedded crown ethers, in which the polarity is unidirectional due to the inner oxygen ring. We note that in experiments, the total charge of the pore will certainly depend on the fabrication methods, the presence of neighboring defects, as well as post-fabrication relaxation. However, for the purposes of our simulations, the total charge of a given multivacancy can be roughly estimated on general chemical grounds, as described below.

An additional advantage of $MoS_2$ pores over graphene-embedded crown ether pores is that the $MoS_2$ pores are predicted to allow the passage of high water flux while rejecting ions,[14-15] desirable in water desalination. Molecular dynamics (MD) simulations of tensile strain applied to a non-ion-conducting pore in $MoS_2$ have shown that it was impermeable to water for small strains and highly permeable to water above a strain of 6% while maintaining robust ion rejection.[16] Given its ability to sustain strains of up to 6 to 11% before breaking,[17] $MoS_2$ overall is a very promising material for strain-controlled permeation in liquid environments.

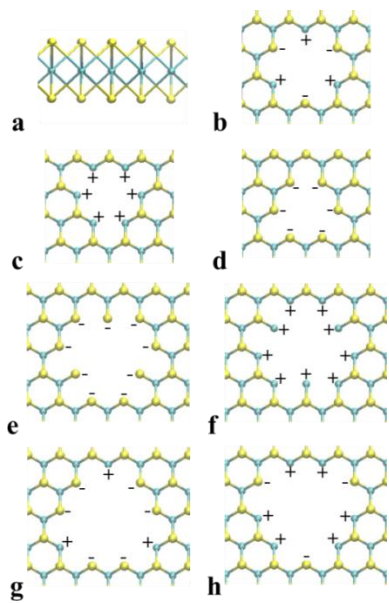

Figure 1. Side-view of the hexagonal MoS$_2$ monolayer (a) and top views of examples of atomically precise sub-nanometer pores with the corresponding expected electrostatic environments at the pore boundaries (b-h). Mo and S atoms are colored teal and yellow, respectively.

Here we propose the use of tensile strain to tune the ion permeation properties of sub-nanoporous single-layer MoS$_2$ membranes. Using MD simulations, we study how electrostatically-driven aqueous ion permeation across sub-nanoporous MoS$_2$ is affected by applied tensile membrane strain. We demonstrate that certain types of pores, which are lined with either entirely Mo or entirely S and thus carry a radial dipole and an effective charge at their edges, exhibit anion or cation conductivity that is highly strain-sensitive. This high degree of mechanosensitivity is shown to occur due to the strain-induced reduction in the overall repulsive ion-pore interactions in an aqueous environment. Depending on the applied strain, membranes containing both cation-conducting and anion-conducting pores can conduct either no ions, only anions, or both anions and cations, providing significant versatility in transport selectivity. This result is especially interesting because the different types of pores discussed here can reside in the same membrane. Unlike in graphene-embedded crown ether pores,[5] no ion trapping occurs within MoS$_2$ pores, due to the effective thickness of MoS$_2$, which prevents ions from achieving a stable partially solvated state while remaining in the center of the pores. We thus demonstrate that highly mechanosensitive ion permeation across solid-state ion channels is a general effect that can result from various mechanisms.

Our results suggest the possibility of actively tuning the conduction of both anions and cations in aqueous solution across nanoporous $MoS_2$ membranes via externally induced tensile lattice strain.

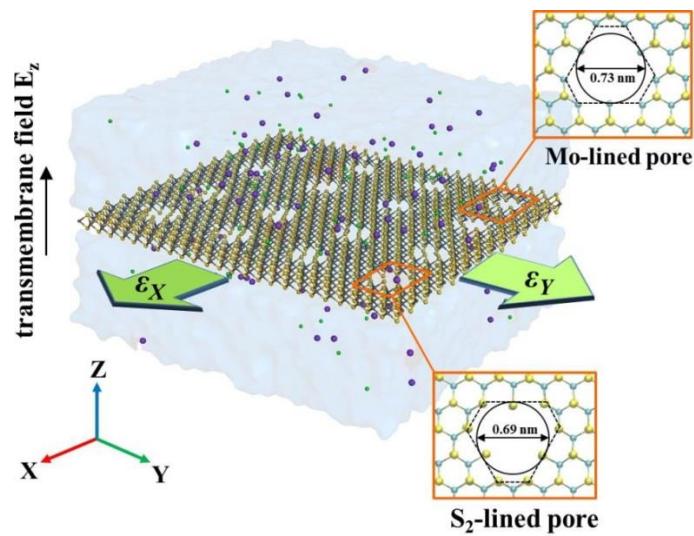

Figure 2. The simulated system: a $MoS_2$ monolayer with 16 pores in aqueous ionic solution. The ions and water are shown as solid spheres and a transparent blue surface, respectively. Eight of the pores correspond to the structure in Fig. 1e (cation-selective $S_2$-lined pore), while the remaining pores are shown in Fig. 1f (anion-selective Mo-lined pore). The dimensions of the system are 8.0 nm × 7.8 nm × 5.0 nm.

**Methods**

The $MoS_2$ model used in our MD simulations was set up according to parameterization described elsewhere[9] and implemented as part of the OPLS-AA forcefield.[18-19] The partial electrostatic charges of Mo and S atoms in the pore regions were calculated using density functional theory (DFT) and refined for use in MD as described in section S1 of the Supporting Information (SI). The DFT calculations were performed using the CP2K package.[20] Water molecules were described by the TIP4P model,[21] and $K^+$, $Na^+$, and $Cl^-$ ions were described using the OPLS-AA forcefield.[18-19]

All MD simulations were performed using the GROMACS v. 2018.1 package.[22-23] Separate systems containing $MoS_2$ membranes in 0.5 M aqueous solutions of NaCl or KCl were assembled as shown in Fig. 2. All simulated systems were pre-relaxed in 5 ns-long semiisotropic (pressure control in the Z-direction, constant simulation box size in the XY-plane) NPT simulations at T = 300 K and p = 0.1 MPa with a 1 fs time step and periodic boundary conditions in all directions. A stochastic velocity-rescaling thermostat and Berendsen barostat were applied with time constants of 0.1 ps and 1 ps, respectively. Production simulations of ion permeation were performed in the NVT ensemble for 100 ns with a 2 fs time step. Ion permeation was induced by applying a constant transmembrane electric field of $E_z = 50$ MV/m. Tensile strain in the membrane was imposed by rescaling the in-plane atomic positions, and edge atoms of each membrane were restrained to their initial positions using springs with a spring constant of 16.61 N/m. Each simulated ionic current value was obtained as the slope of a linear function fitted to the cumulative transmembrane ion flux as a function of simulated time. The corresponding current uncertainties were estimated as described in section S2 of the SI. Ion-pore Gibbs free energies in the form of potential of mean force (PMF) curves were calculated using the Weighted Histogram Analysis Method,[24] as also described in the Methods section of our earlier work.[5]

**Results and Discussion**

We first performed preliminary exploration of the properties of the various types of pores shown in Fig. 1. We observe that the pores depicted in Figs. 1b-d are sterically impermeable to ions, even under significant tensile lattice strain. Note that for the pore shown in Fig. 1c, the observed impermeability to ions is consistent with the results obtained in Ref. 16. On the other hand, the pores shown in Figs. 1g, h exhibit significant permeability at zero applied strain and low mechanosensitivity for Cl$^-$ permeation. Below, we will focus on the two diamond-shaped pore structures shown in Figs. 1e, f, since they were the smallest pores that exhibited weak, yet nonzero, ion permeability at small applied tensile strains, and therefore they are expected to exhibit the greatest mechanosensitivity.

The total charges of the pores, some of which are constructed by removal of not only MoS$_2$ unit cells but also additional Mo or S atoms, are estimated as follows. For ideally ionic or covalent crystals, upon pore fabrication one expects atoms to be ejected in an electrically fully charged or neutral state, respectively. The degree of bond ionicity therefore determines the overall charge that is removed when creating the pore, which one expects to be between the extreme cases of purely ionic and purely covalent crystals. Thus, given the considerable degree of ionicity of bonds in bulk MoS$_2$ monolayers,[9] a straightforward estimate of the total charge of the pore and the surrounding lattice is the opposite of the sum of the bulk partial atomic charges removed to create a given pore structure. Given the bulk partial charges of $q_{Mo}$ = +0.5e and $q_S$ = -0.25e,[9] the structures in Fig. 1e and Fig. 1f carry estimated total charges of -2e and +2e, respectively. Note that in the MD simulations both pore types were introduced in the simulated membrane in equal numbers to enable overall electrical neutrality of the system. It is important to mention that regardless of the total pore charge the radial dipole polarity near the pore edge is expected to be qualitatively intact, as determined by the electronegativities of the constituent atoms. Here, our focus is mainly on the effects of tensile membrane strain on the ion-pore interaction energetics and the transmembrane permeation of aqueous $K^+$, $Na^+$, and $Cl^-$ ions due to their natural abundance and high biophysical relevance.

We expect that the $S_2$-lined pore (Fig. 1e) is impermeable to anions due to the negatively charged pore edge, and potentially permeable to cations, similarly to the graphene-embedded crown ethers.[5, 25] In the same manner, the Mo-lined pore (Fig. 1f) is expected to be impermeable to cations and potentially permeable to anions. The ion-pore interactions in the presence of water are shown in Fig. 3, where we plot the corresponding Gibbs free energy curves as a function of the ion's distance along the Z-direction. The latter is assumed to be the effective reaction coordinate, and the PMFs are presented for the interactions between pores and their counterions, i.e. the positively charged Mo-lined pore interacting with $Cl^-$ and the negatively charged $S_2$-lined pore interacting with $K^+$. A more detailed description of the methodology used to obtain the PMFs in Fig. 3, as well as in our earlier works,[5, 25] is presented in Ref. 24.

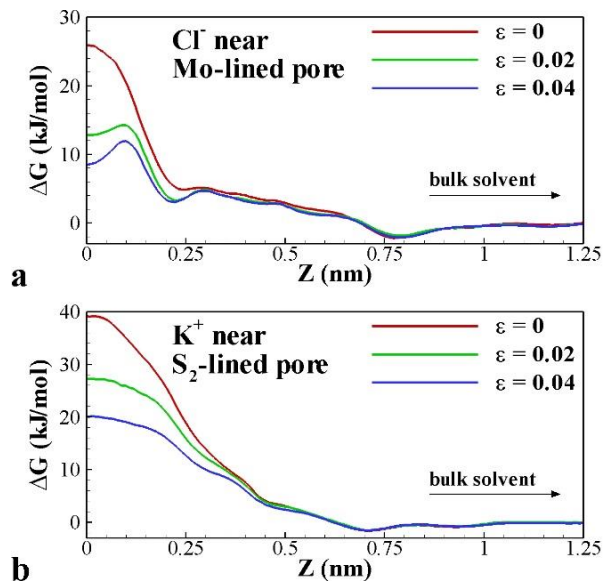

Figure 3. Gibbs free energy in the form of PMF vs. counterion's distance along the Z-axis from the center of (a) a Mo-lined pore (Fig. 1f) and (b) a $S_2$-lined pore (Fig. 1e) for various values of isotropic strain $\varepsilon$. The pores are in the XY-plane (see Fig. 2) and $Z = 0$ corresponds to the vertical position of the sheet formed by the molybdenum atoms within the $MoS_2$ monolayer. The $Z > 0$ half of all profiles (symmetric with respect to the Z-direction) is shown.

In contrast with the ion-trapping graphene-embedded crown ethers, [5, 25] both types of $MoS_2$ pores appear to be overall repulsive to ions. This is due to the combined effect of short-range ion-pore repulsive interactions, as well as ions' dehydration and disruption of the $MoS_2$'s own hydration layer required upon an ion entering the pore. The overall non-trapping nature of the ion-pore interactions observed here may also be partly attributed to the effective thickness of the $MoS_2$ monolayer. In $MoS_2$, which is considerably "thicker" than graphene, the ions are fundamentally unable to maintain a stable partially hydrated state at the center of the pore (along the Z-direction), while interacting with the pore interior. As a result, the energetically preferred location of the ion is on either side of the pore. We note that this dehydration-induced contribution to the observed ions' instability should be reduced in hexagonal boron nitride (hBN) monolayers featuring similar pores,[26] potentially enabling ion trapping.

Compared with the unstrained case, at $\varepsilon = 0.04$ the repulsive barrier $\Delta G_r$ (the difference between the energy peaks and the bulk solvent levels in Fig. 3) is reduced by ~16 kJ/mol and ~19 kJ/mol for the $Cl^-$-Mo-lined pore and $K^+$-$S_2$-lined pore interactions, respectively. With this order of change in energetics, given that the resulting enhancements in permeability are roughly proportional to $\exp\left(\frac{-\delta \Delta G_r}{k_B T}\right) > 100$ (where $k_B$ is the Boltzmann constant and $T = 300$ K is the temperature of the system), the pores essentially on the brink of permeation at $\varepsilon = 0$ are expected to switch to a permeable state with increasing strain. For $K^+$ ions, the concavity of the profile at $Z = 0$ for all shown values of strain suggests instability of the ion near the pore center, enabling rapid ion traversal of the pore confinement. Interestingly, for the $Cl^-$ ion interacting with the Mo-lined pore (Fig. 3a), although the profile is also concave at $\varepsilon = 0$, some convexity and thus a minor local minimum is introduced at $Z = 0$ as strain increases. The latter may cause the ion to traverse the pore confinement more slowly as the pore widens. The likely major underlying cause is a decrease in the short-range repulsive interactions with the Mo atoms at the pore interior. This strain-induced increase in the time spent within the pore is demonstrated later in the text.

The effect of tensile strains on the ionic permeation across $MoS_2$ pores was simulated directly, and the results are shown in Fig. 4a (see Methods for details). The $S_2$-lined pores are found to be impermeable to $Na^+$ and $K^+$ at $\varepsilon = 0$, and a strain-induced onset of permeation is observed. Note that the levels of mechanosensitivity are different for the two cation types, due to the differences in effective ion size and the corresponding solvation effects. The mechanosensitive permeation of $Cl^-$ ions through the Mo-lined pores is considerably higher for the entire range of strain magnitudes, effectively yielding strain ranges that enable membrane permeation of $Cl^-$ only, $Cl^-$ and $Na^+$, and eventually all considered ion species. Compared with the unstrained case, a 28-fold increase in $Cl^-$ current is observed at 4% membrane strain. For $Na^+$ and $K^+$, between $\varepsilon = 0.02$ and $0.04$, permeation is modulated by factors of ~4 and ~6, respectively. Due to the lack of ion trapping in these pores, the permeation process is expected to be weakly competitive here, thus suggesting behavior in salt mixtures similar to that in separate salt solutions, unlike in the competitive permeation of e.g., KCl + NaCl mixtures via graphene-embedded crown ethers.[5]

We also analyzed the times taken by the ions to traverse the immediate pore region (a 0.5-nm-wide region centered on the center plane of the membrane formed by the Mo atoms), and the statistical averages are shown in Fig. 4b. For all ions, pore traversal is rapid, in agreement with our earlier $\Delta G_r$-based observation of ions' instability within the pores. For cations, which traverse the $S_2$-lined pores, these times are nearly strain-independent. For anions moving across the Mo-lined pores, the average transition time is shown to increase with strain. The reported transition times of < 0.1 ns (Fig. 4b) do not constitute Cl⁻ trapping, and the increasing trend is in qualitative agreement with the gradual switch from concavity to convexity observed in the $\Delta G_r$ profile near Z = 0 (Fig. 3a). Note that no significant changes in water accessibility from pore dilation are observed for the Mo-lined pore (see Fig. S3a of the SI), and the effect of membrane strain on anion dynamics within the immediate pore confinement mostly arises from decreasing repulsive interactions with the sterically "large" Mo atoms[9] lining the pore.

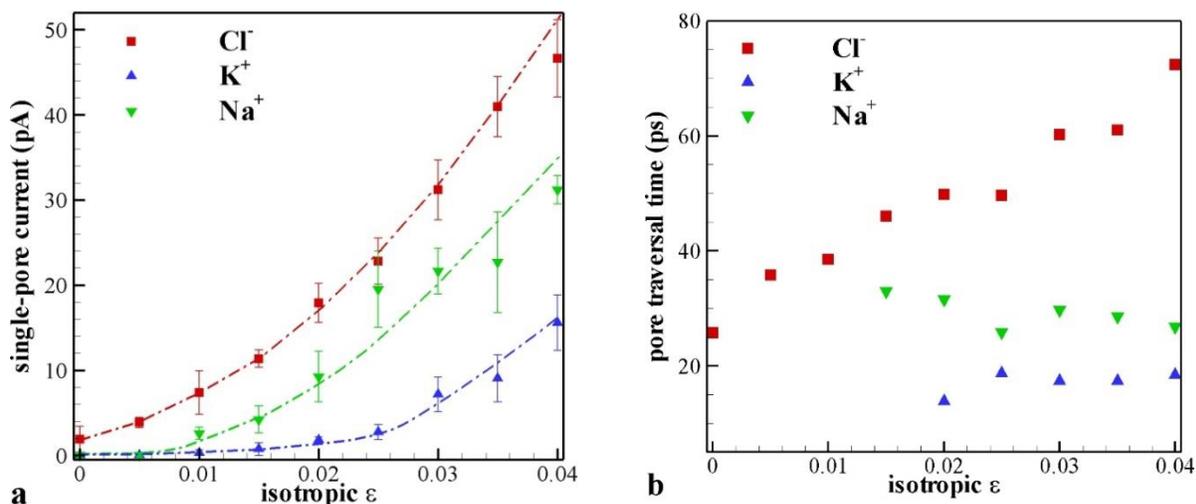

Figure 4. Single-pore ionic currents as functions of isotropic strain ε for water-dissolved KCl and NaCl salts (each salt simulated separately at 0.5 M concentration) (a), as obtained from simulations of the pore array in Fig. 2. Mo-lined pores (Fig. 1f) and $S_2$-lined pores (Fig. 1e) contributed to the vast majority of the anion and cation currents, respectively. The Cl⁻ currents obtained from NaCl and KCl simulations at each value of ε were within 10% of each other so averages between the two simulations are shown. The bars represent uncertainties calculated from fitting different portions of the ionic fluxes (see SI section S2 for details). Dash-dotted lines are visual guides. Average pore region traversal time as a function of ε for all simulated ions, where the pore region is 0.5 nm wide and centered on the center plane of the membrane (b). Traversal times are only reported for the strains with more than ten ion traversals for each ion type.

Overall, we have reported highly mechanosensitive ionic permeation of both cations and anions via selected atomically precise nanopores in $MoS_2$ monolayers. Although the underlying basic mechanisms comprised of the ion-pore and ion-water electrostatic interactions remain the same, strain-gated ion flow across the pores considered here is shown to arise from strain-induced reduction of the overall repulsive ion-pore interactions, unlike in the mechanosensitive graphene-embedded crown ethers,[5] which relied on ion trapping. Our simulations also demonstrate that high mechanosensitivity requires pores with an appropriate effective size and effective charge along the pore lining that results in a pore on the brink of ion permeability, so that the application of tensile lattice strain and subsequent pore dilation leads to significant changes in ion conduction. If a pore is too narrow, it remains impermeable despite the application of strain, while large pores exhibit low relative changes in permeability in response to membrane strain. In the limiting case of pores significantly wider than the characteristic length scales of interatomic interactions, ion conduction would increase by roughly twice the small radial strain, due to increase in the effective pore cross-sectional area.

The versatility of using the same 2D material for introducing a variety of pores, each featuring distinct electrostatic properties and potentially arranged side-by-side, may result in a new class of mechanosensitive nanoporous materials with properties determined by carefully engineered vacancy defects. Similar versatility should be possible in hBN due to the presence of atoms with different electronegativities in the unit cell.[27] Our results qualitatively extend strain-controllable desalination reported earlier[16] to applications of salinity control: here instead of rejecting ions outright, strain tuning of the ion permeation should be possible along with control of the water flow, ranging from near-rejection to significant and selective permeation, depending on the strain magnitude.

The permeation properties of ultra-narrow pores in $MoS_2$ are determined by their geometry and composition, making a discussion of their chemical stability worthwhile. Although pores similar to those considered here have been assumed to be structurally stable in previous modeling work,[14, 16] some degree of pore restructuring is expected for the interior atoms with broken coordination, similarly to the case of narrow pores in hBN.[26] Previous work[28] and our quantum-chemical calculations (see SI) suggest that restructuring indeed takes place without significant effect on the pore structure.

When exposed to oxygen, amorphous and ordered $MoS_2$ oxidizes readily at elevated temperatures of 250-480 °C.[29-31] At room temperature, however, oxidation rates decrease significantly.[30] Furthermore, in aqueous environment, electrochemically produced pores appear to be sufficiently stable to exhibit constant measured ion current baselines as functions of time during DNA translocation.[11] For sub-nanometer pores, stability should be further enhanced because passivation of the narrow pore interiors by e.g., aqueous hydrogens or hydroxyls should be significantly inhibited by the high dehydration barriers of $H^+$ and $OH^-$.[32-33]

**Conclusion**

In summary, we demonstrated the possibility of sensitive strain-gated ionic transport across sub-nanometer $MoS_2$ pores, enabling possible use in applications varying from nanofluidic logical devices to strain-tunable salinity control. No functionalization of the pores beyond defining their geometry via the removal of atoms was required to achieve desired properties. The mechanism underlying the demonstrated high mechanosensitivity, in contrast with the graphene-embedded crown ethers, is shown to boil down to strain-controlled repulsive ion-pore interactions in aqueous environment. Given that the permeation properties of the $MoS_2$ pores presented here in principle depend only on their geometries and bolstered by the fact that similar atomically-precise nanopores have been demonstrated experimentally, our findings suggest the possibility of a new class of mechanosensitive porous materials in which permeation selectivity and mechanosensitivity are determined solely by precise vacancy engineering.

**Supporting Information for Publication**

Supporting Information: Description of partial atomic charges in the pore region and how they were calculated, description of how ion current uncertainties were calculated, and plots of water density profiles along the pore axis.

**Acknowledgements**

We thank A. Noy for suggesting using nanoporous $MoS_2$ membranes for salinity control in systems driven by hydrostatic pressure, A. Davydov for the discussion of multivacancy charge estimation, and R. J. Cannara for the discussion of pore stability in solvent. We are also grateful to A. Govind Rajan and D. Blankschtein for clarifying the details of the bulk $MoS_2$ forcefield they have developed in Ref. 9. Finally, we thank E. Paulechka and A. Kazakov for numerous useful discussions regarding quantum-chemical calculations. This research was performed while A. Fang held a National Research Council (NRC) Postdoctoral Research Associateship at the National Institute of Standards and Technology (NIST). Authors gratefully acknowledge support from the Materials Genome Initiative.

## Disclaimer

**Supporting Information for: "Mechanosensitive Ion Permeation across Sub-Nanoporous MoS$_2$ Monolayers"**

A. Fang, K. Kroenlein, and A. Smolyanitsky*

Applied Chemicals and Materials Division,
National Institute of Standards and Technology
Boulder, CO 80305

Email: alex.smolyanitsky@nist.gov

*S1. Partial atomic charges in the pore region*

All charges were calculated according to the density functional theory (DFT)-based procedure utilized earlier for bulk MoS$_2$ [1]. For our DFT calculations, the total charges of the structures were set to be *-2e* or *+2e* as described in the main text, with a spin multiplicity of 1, corresponding to no dangling bonds. The complete set of partial electrostatic charges obtained from converged DFT calculations and used in the molecular dynamics (MD) simulations is shown in Fig. S1. All atoms <u>without</u> specific charge labels were assigned bulk atomic charges of *+0.5e* and *-0.25e* for Mo and S atoms, respectively [1]. All bonded (harmonic bonds and angles) and non-bonded (Lennard-Jones interactions) parameters were assigned according to the bulk MoS$_2$ parameterization [1], with the exception of the atoms selected in black rectangles in the S$_2$-lined pore in Fig. S1. For these atoms, to achieve a better agreement between the relaxed pore structures obtained from DFT calculations and their MD counterparts, the $\theta_0$ parameter for the corresponding S-Mo-S angle was reduced from 83.8 to 37.0 degrees due to S-S bond formation.

We note that in the pore regions the DFT calculations suggest that the perturbation of the electron densities relative to those in bulk MoS$_2$ extends beyond the first two in-plane atomic layers nearest to the pore edge (pore edge atoms and their nearest neighbors). To simplify charge assignment and thus facilitate reproducibility of our results, we effectively "condensed" the perturbations to the shown region so that the individual charges remained close to the "raw" DFT-calculated charges, while the total electrostatic potential experienced by a test ion in the center of each pore is within 10 % of that obtained from the DFT-based charges. The reader is encouraged to verify that the sums of labeled charges in Fig. S1 correspond to those obtained from bulk-charge assignment, as required by charge conservation. With bulk charge assignment, for the Mo-lined pore, the sum in the two-layer region is *9 × 0.5e – 24 × 0.25e = -1.5e*. For the S$_2$-lined pore, the sum is *-18 × 0.25e + 12 × 0.5e = +1.5e*. The total pore charges are expectedly $q_{Mo-lined}$ = *+2e* and $q_{S_2-lined}$ = *-2e*.

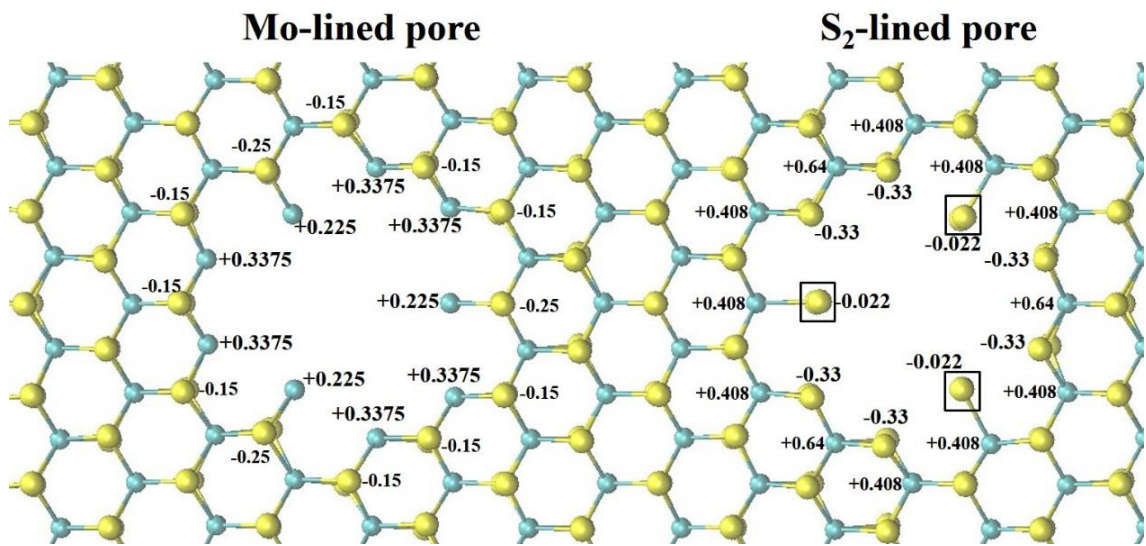

Figure S1. Distribution of partial atomic charges in the pore regions for the Mo-lined (left) and $S_2$-lined (right) pores in $MoS_2$ monolayer (Mo and S atoms are colored teal and yellow, respectively). All values are shown in the units of elementary charge.

## S2. Uncertainty in simulated permeation

Shown in Fig. S2 is an example of ionic fluxes obtained for 0.5 M NaCl at $\varepsilon = 0.025$. The current for each ion type within each bin is calculated from the slope of the corresponding portion of the flux and the uncertainty is calculated as the standard deviation of the slopes obtained from all bins. Two bins were used for the data shown in Fig. 4a.

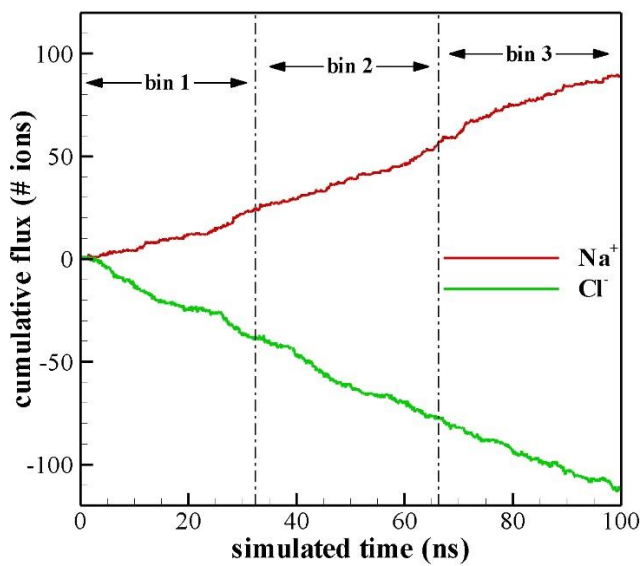

Figure S2. Cumulative ionic fluxes with time binning to estimate the uncertainty due to flux variations.

*S3. Water distribution along the pore axis*

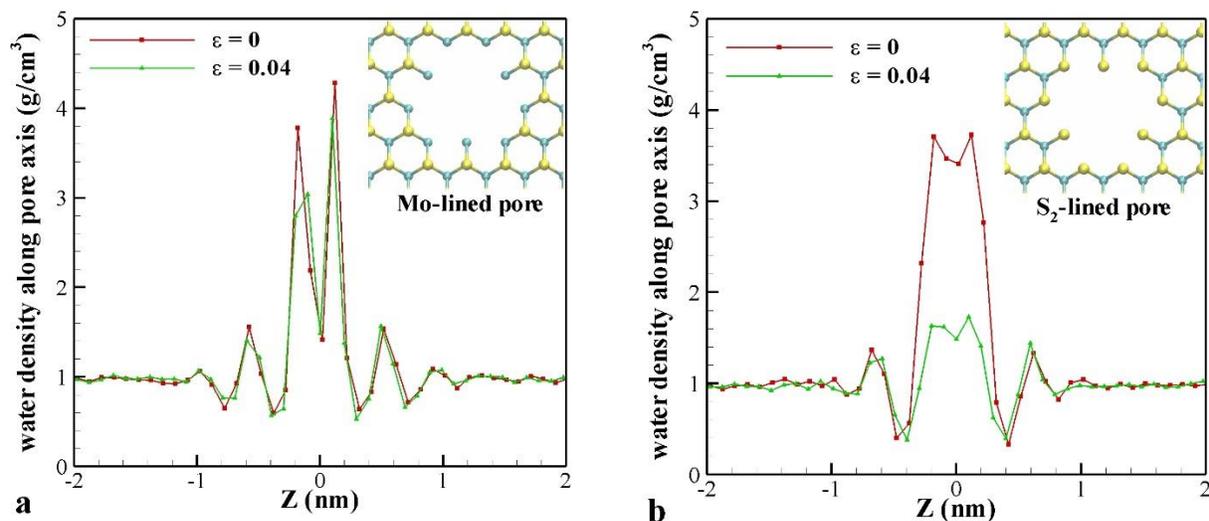

Figure S3. Water density along an axis normal to the membrane and passing through the center of the Mo-lined (a) and $S_2$-lined (b) pore for the unstrained case and $\varepsilon = 0.04$. The pores are in the XY-plane and $Z = 0$ corresponds to the vertical position of the sheet formed by the molybdenum atoms within the $MoS_2$ monolayer. The water density profiles shown here are interpolated from a three-dimensional water density distribution calculated from water oxygen positions.